\documentclass[referee,a4paper,12pt,traditabstract]{swsc} 

\usepackage{graphicx}
\usepackage{txfonts}
\usepackage{caption}
\usepackage{subcaption}
\usepackage{placeins}
\usepackage{epstopdf}
\usepackage[mathlines]{lineno}
\usepackage[authoryear,round]{natbib}
\usepackage[backref]{hyperref}
\usepackage{url}
\usepackage{xcolor}
\usepackage{acronym}
\acrodef{SOHO}[SOHO]{SOlar and Heliospheric Observatory}
\acrodef{EPHIN}[EPHIN]{{Electron Proton Helium Instrument}}
\acrodef{SEP}[SEP]{{Solar energetic particle}}
\acrodef{CME}[CME]{Coronal Mass Ejection}
\acrodef{GCR}[GCR]{Galactic Cosmic Ray}
\acrodef{FD}[FD]{Forbush Decrease}
\acrodef{SSD}[SSD]{Solid State Detector}
\acrodef{ADC}[ADC]{Analog Digital Conversion}
\acrodef{GEANT4}[GEANT4]{GEometry ANd Tracking}
\acrodef{PHA}[PHA]{Pulse Height Analysis}
\acrodef{VDA}[VDA]{velocity dispersion analysis}

\hypersetup{colorlinks=true,citecolor=cyan,urlcolor=cyan,linkcolor=blue}

\begin{document}

   \title{The Electron Proton Helium INstrument as an Example for a Space Weather Radiation Instrument}

   \titlerunning{EPHIN as space weather instrument}

   \authorrunning{K\"uhl et al.}

   \author{Patrick Kühl\inst{1}, Bernd Heber\inst{1}, Ra\'ul G\'omez-Herrero\inst{2}, Olga Malandraki\inst{3}, Arik Posner\inst{4}, Holger Sierks \inst{5}}

   \institute{Institut f\"ur Experimentelle und Angewandte Physik, Universit\"at Kiel, Kiel, Germany
              \email{\href{mailto:kuehl@physik.uni-kiel.de}{kuehl@physik.uni-kiel.de}}
        \and
        Universidad de Alcal\'a, Space Research Group, 28805 Alcal\'a de Henares, Spain
        \and
        National Observatory of Athens/IAASARS, Athens, Greece
        \and
        SMD/Heliophysics Division, NASA HQ, Washington, DC, USA,
        \and 
        Max-Planck-Institut für Sonnensystemforschung, G{\"o}ttingen, Germany.
             }

   \date{\today}

 
  \abstract
   {The near-Earth energetic particle environment has been monitored since the 1970's. With the increasing importance of quantifying the radiation risk for, e.g. for the human exploration of the Moon and Mars, it is essential to continue and further improve these measurements. The Electron Proton Helium INstrument (EPHIN) on-board SOHO continually provides these data sets to the solar science and space weather communities since 1995. Here, we introduce the numerous data products developed over the years and present space weather related applications. Important design features that have led to EPHINs success as well as lessons learned and possible improvements to the instrument are also discussed with respect to the next generation of particle detectors.}       
   

   \maketitle

\section{Introduction}
\label{sec:ephin}
\acp{SEP}, from suprathermal (few keV) up to relativistic (few GeV for protons and ions) energies are an important contributor to the space environment characterization. They are emitted from the Sun in association with solar flares and \ac{CME}-driven shock waves. \ac{SEP} events constitute a serious radiation hazard and pose a threat to modern technology which relies on spacecraft as well as to humans in space. In addition they are of concern for avionics and commercial aviation. Thus, mitigation procedures must be developed. Novel \ac{SEP} event forecasting tools have been developed within the HESPERIA H2020 EU project and are strongly relied upon to mitigate against \ac{SEP} events.\newline
\begin{figure}
\begin{center}
\includegraphics[width=1\columnwidth]{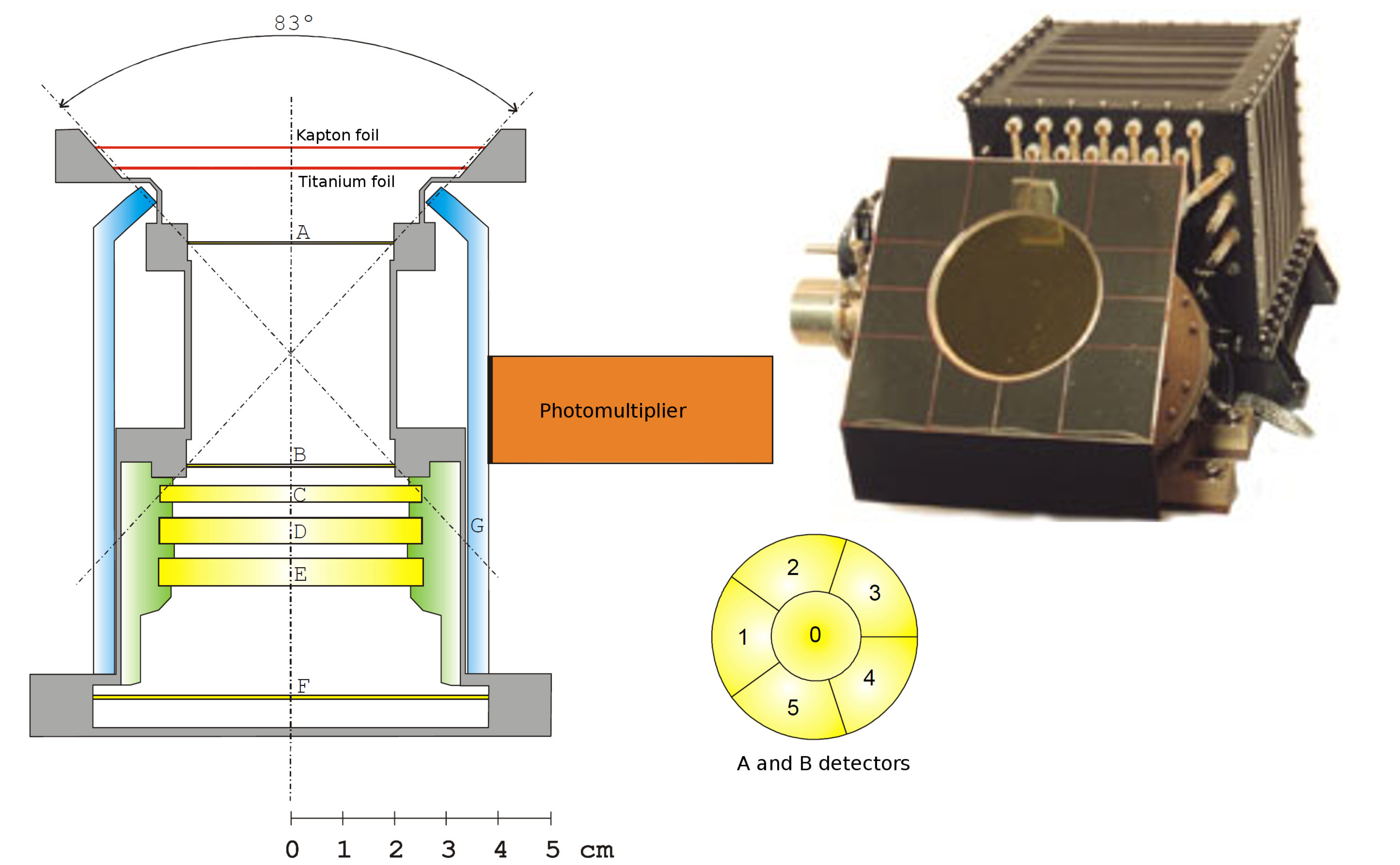} 
\end{center}
\caption{Sketch (left) and photograph (right) of the EPHIN instrument.}
\label{fig:EPHIN}
\end{figure}
These forecasting tools as well as the scientific studies of the very events they are designed for to forecast, naturally share some limitations such as the availability and quality of the underlying data. Arguably one of the most import data sources for space weather applications is the NASA/ESA \ac{SOHO} that was launched in 1995 and is since 1996 orbiting the Lagrangian point L1. The scientific payload of the spacecraft consists of several remote and in-situ instruments including the \ac{EPHIN}, a particle telescope with a field of view of about 83$^\circ$ and a geometry factor of 5.1 cm$^2$~sr that measures electrons with energies between 0.25 and 10.4 MeV as well as protons and Helium in the energy range of 4.3 up to above 53 MeV/nucleon \citep{Mueller-Mellin-etal-1995}. A schematic view of the EPHIN instrument is shown in Figure \ref{fig:EPHIN}. The instrument consists of a stack of six \acp{SSD} (labeled A-F in Fig.~\ref{fig:EPHIN}) that are surrounded with a scintillation detector (G), that acts as an anti-coincidence. Stopping ions are identified by applying the $\frac{dE}{dx}-E$-method. Since the telemetry data rate is limited, energy losses in \acp{SSD} A-E can only be transmitted for a statistical ensemble of all particles counted by \ac{EPHIN}. This constitutes the so-called \ac{PHA} data. A crude onboard separation between electrons, protons and helium has been introduced for different penetration depths, i.e. different coincidences, by comparing the energy-loss in \ac{SSD} A to different thresholds. Futhermore, \acp{SSD} A and B are segmented (Fig. \ref{fig:EPHIN}) with the outer segments being switched-off automatically by the instrument during periods of high fluxes (i.e. if the count rate of the central segment of detector A is above 15000 counts per second) in order to reduce the geometric factor by a factor of 24 resulting in fewer electronic pile-up and dead time issues \citep[for details see][]{Mueller-Mellin-etal-1995}.\newline
The instrument has been proven to be of significant value not only for the space science but also for the space weather community especially due to its consistent measurements over two and a half decades as well as its high alive time during the mission. Certain long-term measurements such as those of MeV electrons with a low background are even unique to EPHIN. Apart from the communication loss to SOHO in 1998, \ac{EPHIN} has a data coverage of over 90\%. Figure \ref{fig:lifetime} clearly shows that most of the data gaps correspond with roll maneuvers of the SOHO spacecraft (indicated by blue lines). The reason for the SOHO roll maneuvers are mechanical issues with SOHO's high gain antenna starting in 2003 \citep{soho_keyhole}. In order to reduce the limitations in telemetry that come with an antenna unable to actively point to Earth, it was decided to move the antenna one last time to a "sweet spot" that could guarantee a stable connection to Earth during half of SOHOs orbit around L1 and rotate the entire SOHO spacecraft twice per orbit (i.e. every three months). In-between both half-orbits (around the time of the roll maneuver), however, a short period during which communication is limited remains. These keyhole periods are the reason for the majority of EPHIN's data gaps. Note that the pointing of EPHIN (nominal: 45 deg west from sun-s/c line, i.e. along the nominal Parker spiral) is also affected by the roll maneuvers (pointing becomes perpendicular to the nominal spiral). The dates of the roll maneuvers as well as the attitude of spacecraft can be found here: \url{https://soho.nascom.nasa.gov/data/ancillary/attitude/roll/nominal_roll_attitude.dat}.
The keyhole periods are also listed here: \url{https://sohowww.nascom.nasa.gov/soc/keyholes.txt}\newline
Despite being active and measuring particles for almost 25 years - the primary mission duration was defined as two years \citep{soho} -, forced inactivity and possible harsh temperature environments for three months during the communication loss of SOHO in 1998 \citep{soho_recovery} as well as being exposed to various large \ac{SEP} events \citep[][and references therein]{kuehl_etal_2019}, EPHIN remains in a remarkably good condition and is still providing data for and fulfilling most of its primary scientific objectives. The negative impact of the only two technical issues, increased noise level 1) in detector E in 1997 and in 2) in detector D in 2017, was limited by careful software management and data analysis. After the noise increases in the detectors were monitored, EPHIN was commanded to failure modes "FME" and "FMD" on February, 19th, 1997 and October, 4th, 2017, respectively. The failure modes remove the defective detectors from the coincidence logic which results in lower number of energy channels per particle type (four in nominal mode, three and two for FME and FMD, respectively). It was possible, however, to retrieve the nominal energy channels based on the energy losses in the \acp{SSD} A, B and C using \ac{PHA} data (see section \ref{sec:level3}).\newline
Section~\ref{sec:dataproducts} provides an overview of all different data products of EPHIN including advantages and disadvantages as well as applications of the different data products. Section~\ref{sec:spaceweatherapplications} describes applications of the EPHIN data with respect to space weather such as investigations and forecasts of \ac{SEP} events, in section~\ref{sec:lessonslearned} limitations and possible improvements for EPHIN are discussed based on the lessons learned. A summary concludes the importance of measurements from EPHIN for the space weather community and proposes design features for future instruments.\newline

\begin{figure}
    \centering
    \includegraphics[width=1\columnwidth]{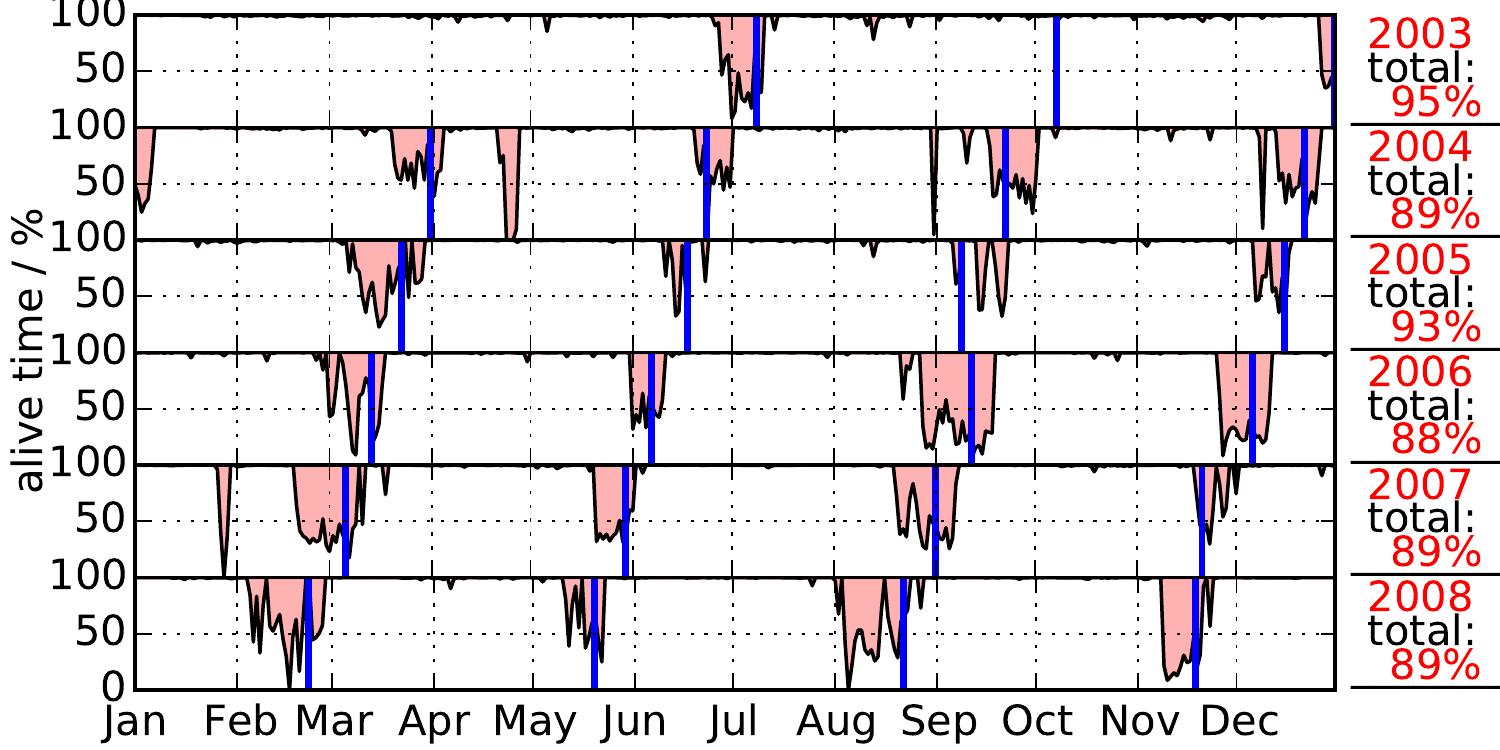}
    \caption{Alive time of EPHIN over six years. The blue lines indicate the date of a SOHO roll maneuver.}
    \label{fig:lifetime}
\end{figure}{}

\section{EPHIN Data Products}
\label{sec:dataproducts}
The raw (level 0) EPHIN data telemetered to ground is processed and formatted routinely at the Christian-Albrechts University of Kiel. Several data levels and file formats have been implemented and released publicly over the years. All data products are provided as merged daily files with a cadence of 1~minute except for the histogram data (see section \ref{sec:histogram}), which has a time resolution of eight minutes. The following section introduces the most important data products, flaws and advantages, download sources for the data as well as their usage for space weather applications. While the majority of these data sets are available only with a couple of days of delay after acquisition onboard, some limited data are also available in near real-time (see section \ref{sec:nearrealtime}). Documentations of the different data products and their file format can be found here: \url{http://ulysses.physik.uni-kiel.de/costep/doc/}

\subsection{Level~1}
\label{sec:level1}
The Level~1 data products consist of 1) housekeeping data (not covered here), 2) single \& coincidence counter as well as 3) histograms, and 4) \ac{PHA} data. The SCI files that include the counters and histograms as well as the \ac{PHA} files can be downloaded from \url{http://ulysses.physik.uni-kiel.de/costep/level1/}

\subsubsection{Single Counter Rates}
\begin{figure}
    \centering
    \includegraphics[width=1\columnwidth]{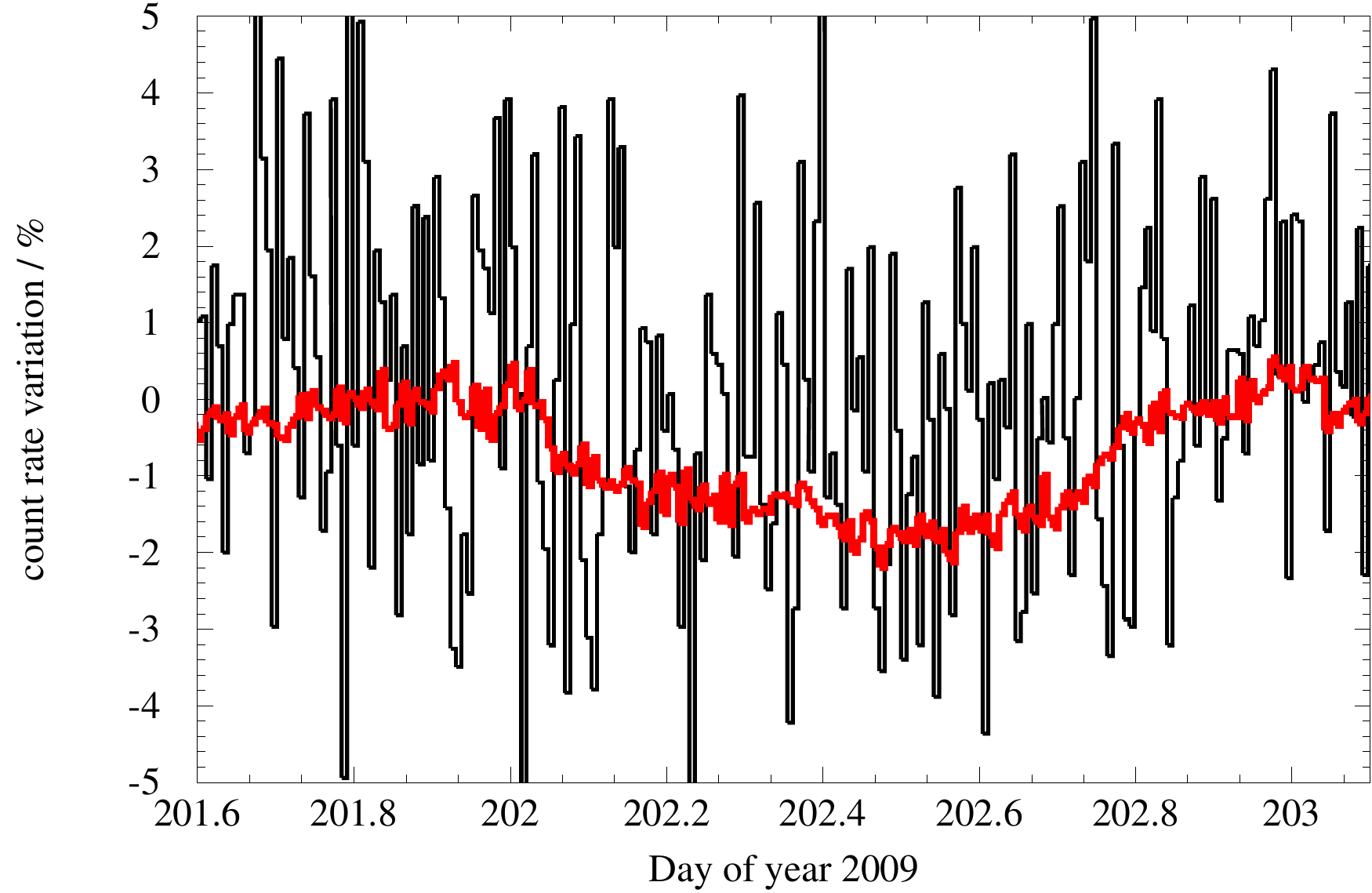}
    \caption{Variation of 10 minute averaged count rates of the single detector rate G (red curve) and the integral channel (black curve) during the passage of a Stealth \ac{CME} \citep{heber_etal_2015}. For details see text.}
    \label{fig:fd}
\end{figure}

\begin{figure}
    \centering
    \includegraphics[width=0.9\columnwidth]{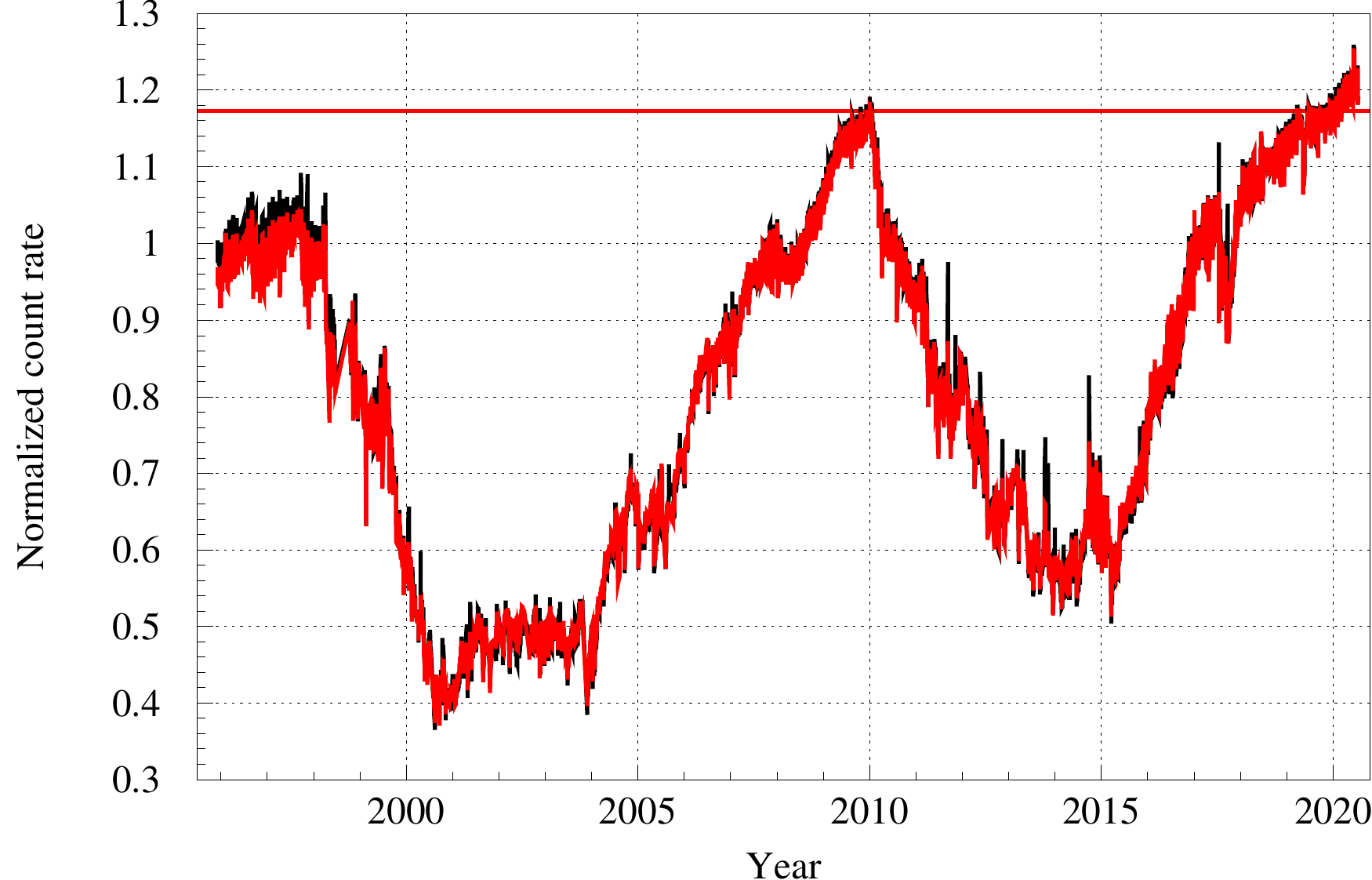} \\
    \includegraphics[width=0.9\columnwidth]{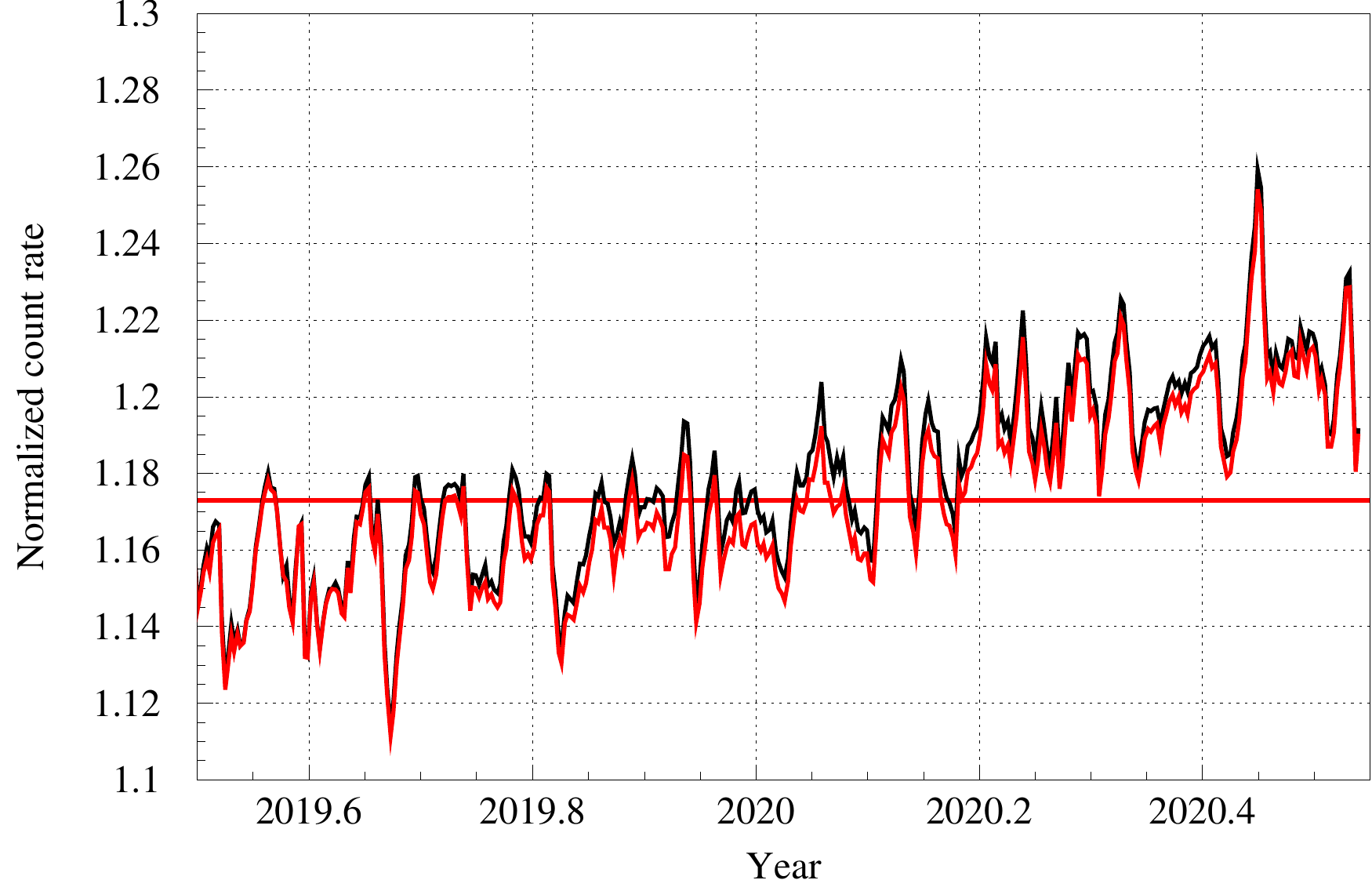}
    \caption{Uper and lower panel display the daily averaged quiet time galactic cosmic ray flux variation over the last two solar cycles and from July 2019 to July 2020, respectively. The black and red line show the count rate variation of the anti coincidence detector G and the \ac{SSD} F using counting rate thresholds in \ac{SSD} B omitting time periods of increased energetic particle fluxes (for details see text).}
    \label{fig:GCR-Modulation}
\end{figure}

\begin{figure}
    \centering
    \includegraphics[width=1\columnwidth]{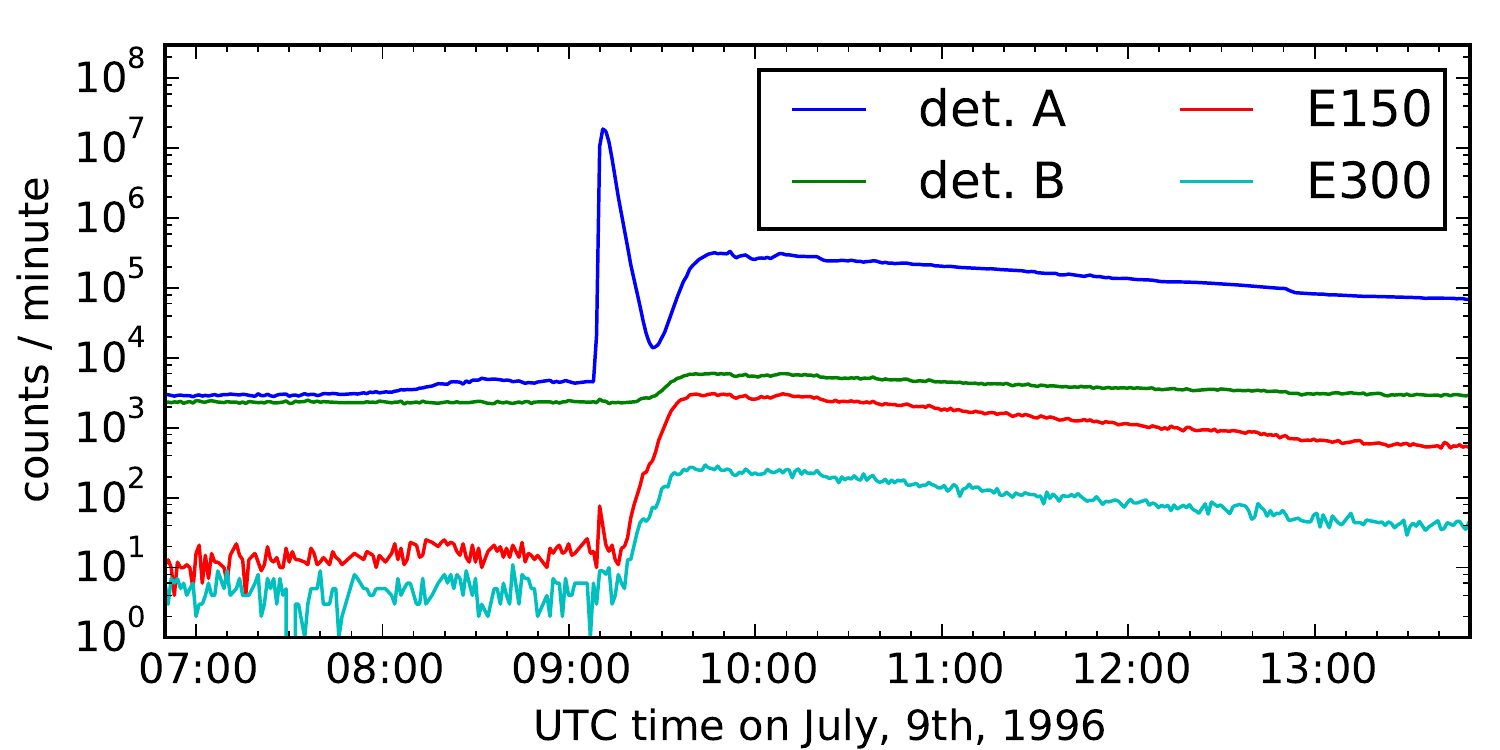}
    \caption{One minute resolution count rates of the electron channels E150 (red), E300 (teal), the single detectors A (blue) and B (green) on July 9, 1996. }
    \label{fig:singlecounter}
\end{figure}

Independent of any coincidence logic, EPHIN tracks the count rate of each individual detector (including the anticoincidence G). Whether or not a particle is counted in a detector depends on its energy deposition in the detector in comparison to the detectors lower threshold \citep[energy thresholds are given in table 3,][]{Mueller-Mellin-etal-1995}. While originally included in the data stream in order to monitor the noise level of the detectors, it has been shown that these single counter rates are also of scientific value due to their high statistics \citep[an idea first proposed by][]{richardson_etal_1996}. \citet{heber_etal_2015} has shown that \ac{EPHIN} is capable of measuring count rate variations with amplitudes in the below 1\% regime on an hourly time resolution and hence, observe \acp{FD} associated to ICMEs in general and stealth \acp{CME} in particular \citep[see for example][]{Howard-and-Harrison-2013}. Utilizing the  anti-coincidence detector G the measurement capabilities allow to investigate \acp{FD} using accumulation times as low as 10 minutes as shown in Figure~\ref{fig:fd}. 
The figure shows a time period discussed in detail in \cite{heber_etal_2015} when a \acf{FD} caused by Stealth \ac{CME} was observed. It compares the count rate variations of the single detector G (red curve) to that of the integral channel (i.e. coincidence of \acp{SSD} A to F, black curve). Both data sets have been averaged over 10 minutes and are normalized to the count rate measured during the time period from 19:00 to 22:30 UT on July 20, 2009. While the count rate for the single detector is about 27,000 counts/minute the integral channel counts about 240 counts/minute leading to a 10 times higher statistical accuracy of the single detector compared to the integral count rate. The corresponding statistical uncertainty of the measurements shown in Figure~\ref{fig:fd} are 0.19\% and 2.0\%, respectively. An additional systematic error results from the resources of the \ac{ADC} that can be approximated to be of the order of 0.15\% for detector G resulting in a total uncertainty of 0.25\%. Thus variations on a 0.5\% and 1\% level have a significance of 2 and 4$\sigma$, respectively. From the figure it is clear, that single detectors with a count rate of $\approx$300,000 counts per 10 minutes can resolve cosmic ray variations that are as small as one percent in less than an hour while the nominal coincidence channels (hundreds of counts per minute) do not offer sufficient statistics for such an analysis. The disadvantage of the single counter on the other side is the loss of any direct information on the particle types and energies responsible for the count rate. However, by calculating the energy dependent response for each single counter, this information can be indirectly derived to some extend allowing the counters for example to be used to study \ac{GCR} modulation \citep{kuehl_etal_2015_b}. Figure~\ref{fig:GCR-Modulation} displays the daily averaged quiet time \ac{GCR} flux variation over the last two solar cycles and from July 2019 to July 2020, respectively. The black and red line show the count rate variation of the anti coincidence detector G and the \ac{SSD} F, respectively. A count rate threshold in \ac{SSD} B was used to omit time periods of increased energetic particle fluxes (e.g. \acp{SEP}). Although both curves follow each other during most times the anti-coincidence detector G reflects some increases not measured by the \ac{SSD} F. The reason for this lies in the response of G to low energy electrons that are not measured by \ac{SSD} B. The red line reflects level of the highest galactic cosmic ray flux ever reported during the space era \citep{Zhao-etal-2014}. From the lower panel of the figure it is evident, that we reached a higher flux level in 2020 leading to a harsher \ac{GCR} caused radiation environment close to Earth than in the previous solar minimum. 
\newline
In addition to particles, photons from solar flares can also cause energy depositions in the detectors. Especially detector A with a threshold of only 30~keV is sensitive to X-ray. Figure \ref{fig:singlecounter} shows the count rate of the lower two electron channels (red and teal) as well as the single counter rate of detector A (blue) and B (green) during the \ac{SEP} event on July 9, 1996 \citep[see for example][]{Droege-etal-1997}. All channels and single counter show an increase in count rate starting at 9:20~UT with a maximum around 9:40~UT followed by a gradual decrease due to \acp{SEP}. However, the detector A single count rate (as well as the lowest electron channel up to small extent) have another maximum at approximately 9:10~UT lasting only for a couple of minutes that has been caused by photons interacting in the detector \citep[compare to ][and references therein]{posner_2007}. The photons stem from an X2.6 x-ray flare located at S10 and W30 that had an onset, maximum and end time at 09:01 UT, 09:12 UT and 09:49 UT, respectively (from ftp://ftp.ngdc.noaa.gov/STP/space-weather/solar-data/solar-features/solar-flares/x-rays/goes/xrs/). These capabilities of detecting solar flare photons can be exploited for studies on and/or forecasting schemes \citep[see for example][and references therein]{Nunez-etal-2018}. It has to be noted, however, that the flare signal presented in Figure \ref{fig:singlecounter} was measured during an early phase of the mission. Since the background noise of detector A increased from roughly 1000 counts per minute to the current level of 1-3~10$^5$ counts per minute over the first ten years of the mission, the sensitivity of EPHIN to solar flare photons is restricted to larger flares (M and X class) for the majority of the mission.
{}

\subsubsection{Histograms}
\label{sec:histogram}
EPHIN provides histograms of the total energy deposition in the detector stack. These histograms are given for four different particle penetration depths (e.g. particles stopping in B, C, D or E) with 64 bins each (corresponding to roughly 30 channels populated by proton and helium each in total). Although the energy deposition is calculated using the \ac{PHA} electronic chain on-board, the histogram data is not to be confused with the \ac{PHA} data product which is an entirely independent data product. The histograms are averaged over eight minutes due to bandwidth limitations. Furthermore, it is not possible to distinguish between different isotopes (i.e. $^3$He/$^4$He) based on the histograms. However, the histogram data combines most advantages from \ac{PHA} data (higher energy resolution than coincidence channels, better particle identification) with the good statistics from the counter data products (in contrast to \ac{PHA}, which only represents a statistical sample). Hence, applications such as a \ac{VDA} during the onset of an \ac{SEP} event \citep[e.g.][]{krucker_lin_2000} can benefit from this data set. While the time-resolution of 8~minutes is relatively large for a \ac{VDA}, EPHINs large energy range from 4.3 to 53~MeV allows to measure protons with differences in arrival time of roughly 80~minutes (assuming a 1.3~AU parker spiral) and hence, still provide meaningfull \ac{VDA} results \citep[for a \ac{VDA} based on EPHIN data see e.g.][]{kleinposner}. Furthermore, the calculation of fluence spectra integrated over an entire \ac{SEP} event \citep[e.g.][]{mewaldt_etal_2005} is possible based on this data set. While the Level~1 histograms only provide counts at the different energy deposition, a new data product providing fluxes and sophisticated particle separation is under preparation.

\subsubsection{Coincidence Counter}
\label{sec:coincidence}
The coincidence counters contain counting rates for individual particle species plus an additional channel for penetrating particles (the integral channel). For the four ranges stopping in the detector stack (particle stopping either in B, C, D or E), different thresholds for detector A \citep[given in table 3,][]{Mueller-Mellin-etal-1995} are used to distinguish roughly onboard between electrons, protons and helium particles (resulting in 12 stopping channels in total). Note that particles heavier than helium can not be resolved by EPHIN due to the usage of linear amplifiers in the electronics and hence, they will be counted in the helium channel. Since the contribution of these heavier particles can be considered small with respect to the helium flux \citep[see][for SEPs and \acp{GCR}, respectively]{reames_2019,george_etal_2009}  the systematic uncertainty introduced by this false identification can be neglected. The simple particle identification by the thresholds, however, has significant flaws regarding the electron, proton and helium identification as well. Electrons can spoil the proton channel, while the helium and electron channels also have a finite response to protons. These small relative contributions can get significant depending on the measured actual flux environment (e.g. electron rich \ac{SEP} events, or \ac{SEP} events with a small helium to proton ratio). The advantage of this data product is the higher statistics in comparison to other data products with improved particle separation such as the Level~3 data (section \ref{sec:level3}).\newline
For the proton and helium measurements, the count rates are further divided according to trajectory inclination relative to the sensor axis. This information can be used for path-length corrections of the energy depositions resulting in an improved particle separation (i.e. isotopic resolution).

\subsubsection{Level~1 PHA data}
The Level~1 \acf{PHA} contains detailed information for a statistical sample (a simple priority table is ensuring at least a couple of helium particles) of the measured particles. For those particles covered, the segments triggered in detectors A and B, the identified coincidence, as well as the height of the voltage pulse measured by the EPHIN electronics that was caused by the energy deposition of the particle and the used amplifier (low or high gain) is given for each detector \citep[for details see][]{Mueller-Mellin-etal-1995}. Hence, in contrast to the histogram data, not only the total energy deposition of a particle in the detector stack but also the information regarding the variations and changes in energy deposition along its trajectory through the various detectors is available. However, we strongly recommend to utilize the Level-2 \ac{PHA} data sets as explained below.

\subsection{Level~2}
The Level~2 data consists of 1) flux data for electrons, protons and helium particles with four energy channels each as well as one penetrating channel (RL2) data as well as 2) \ac{PHA} data in units of MeV. Both data sets can be downloaded from \url{http://ulysses.physik.uni-kiel.de/costep/level2/}
\subsubsection{RL2 data}
Based on the Level~1 coincidence counters and simulations of the instrument yielding the response function of the instrument for different particles and energies, the flux is calculated for all twelve stopping channels (i.e. four channels for electrons, protons and helium) in units of (cm$^2$~sr~s~MeV/nucleon)$^{-1}$ plus the integral channel counting particles producing signals in detectors A-F in units of (cm$^2$~sr~s)$^{-1}$. While the application of such response factors is straightforward for ions due to their almost constant response over energy \citep[see nominal proton response function in figure A1 in the appendix from][]{kuehl_etal_2019}, the large amount of scattering of electrons in the detector results in a more complex energy response typically requiring a more detailed treatment such as the bow-tie analysis \citep[and references therein]{bowtie}. For the RL2 files, such an analysis was not performed and the response has been only approximated by integration over the energy dependent response functions resulting in systematic uncertainties which are depending on the spectral shape of the observed electrons. Other advantages and disadvantages of this data product include those given in section \ref{sec:coincidence}.

\subsubsection{Level~2 PHA data}
\label{sec:lvl2pha}
Based on the Level~1 \acf{PHA} and the calibration of the instrument, the energy deposition in each detector can be calculated. The Level~2 \ac{PHA} contains this detailed information for a statistical sample (a simple priority table is ensuring at least a couple of helium particles) of the measured particles. For those particles covered, the segments triggered in detectors A and B, the identified coincidence, as well as energy deposition in units of MeV of the particle in each detectors given. \newline
Utilizing the Level~2 \ac{PHA} data, EPHIN is capable of providing energy spectra with small \& variable energy channels, identification and separation of different isotopes (i.e. $^3$He/$^4$He) as well as more complex, new data products (see section \ref{sec:level3} and \ref{sec:penetrating}).\newline
As mentioned above the first two \acp{SSD} are sectorized in order to determine the different path length of particles in order to improve the determination of the $^3$He/$^4$He ratio \citep{Mueller-Mellin-etal-1995}. The \ac{PHA} files store more detailed information, i.e. which sector is hit in the first two \acp{SSD} for every single particle evaluated. \citet{banjac_etal_2015} has shown that 36 virtual telescopes defined by the different sector combinations can be utilized to investigate the particle pitch angle distribution in the field of view of the instrument with an approach recently also proposed for the Parker Solar Probe's IS\sun IS/EPI-Hi \citep{mccomas_etal_2016}. Unfortunately, the counting statistics of the \ac{PHA} data only allows to determine the deviation from an isotropic distribution for only a few events. 

\subsection{Level~3 Ion Fluxes}
\label{sec:level3}
Due to increasing noise in detector D, EPHIN was commanded to Failure mode D in October, 2017, (see section \ref{sec:ephin}) limiting the energy resolution of the nominal count rate based data products. Therefore, \citet{kuehl_etal_2019} have developed a method based on \ac{PHA} data in order to restore the original energy resolution for protons and helium using only the information of energy depositions in detectors A, B and C. The data product has been successfully validated against ACE/SIS, SOHO/ERNE and GOES11/EPS. However, due to the methodology, neither electron fluxes nor isotopic resolution (i.e. $^3$He/$^4$He) could be retrieved by this method. Furthermore, the \ac{PHA} based nature of the data product can cause time-dependent variations of the statistical uncertainty (i.e. at the onset of an \ac{SEP} event, when electrons are filling most of the \ac{PHA} buffer). Nonetheless, the long-term availability of the measurements make the data especially interesting for space weather studies and monitoring \citep[cf. ][]{kuehl_etal_2019}. The resulting data product offers consistent measurements from 1995 on, various time resolutions (from 1 minute to daily averages) as well as statistical and systematical uncertainty in a user friendly file format accessible here: \url{http://ulysses.physik.uni-kiel.de/costep/level3/l3i/}. Furthermore, the method has been extended such that the flux of protons and helium nuclei can be derived for any energy channel in the range from 4 to 53~MeV/nucleon allowing an easy inter-calibration between various different missions.

\subsection{Penetrating particles}
\begin{figure}
    \centering
    \includegraphics[width=1\columnwidth]{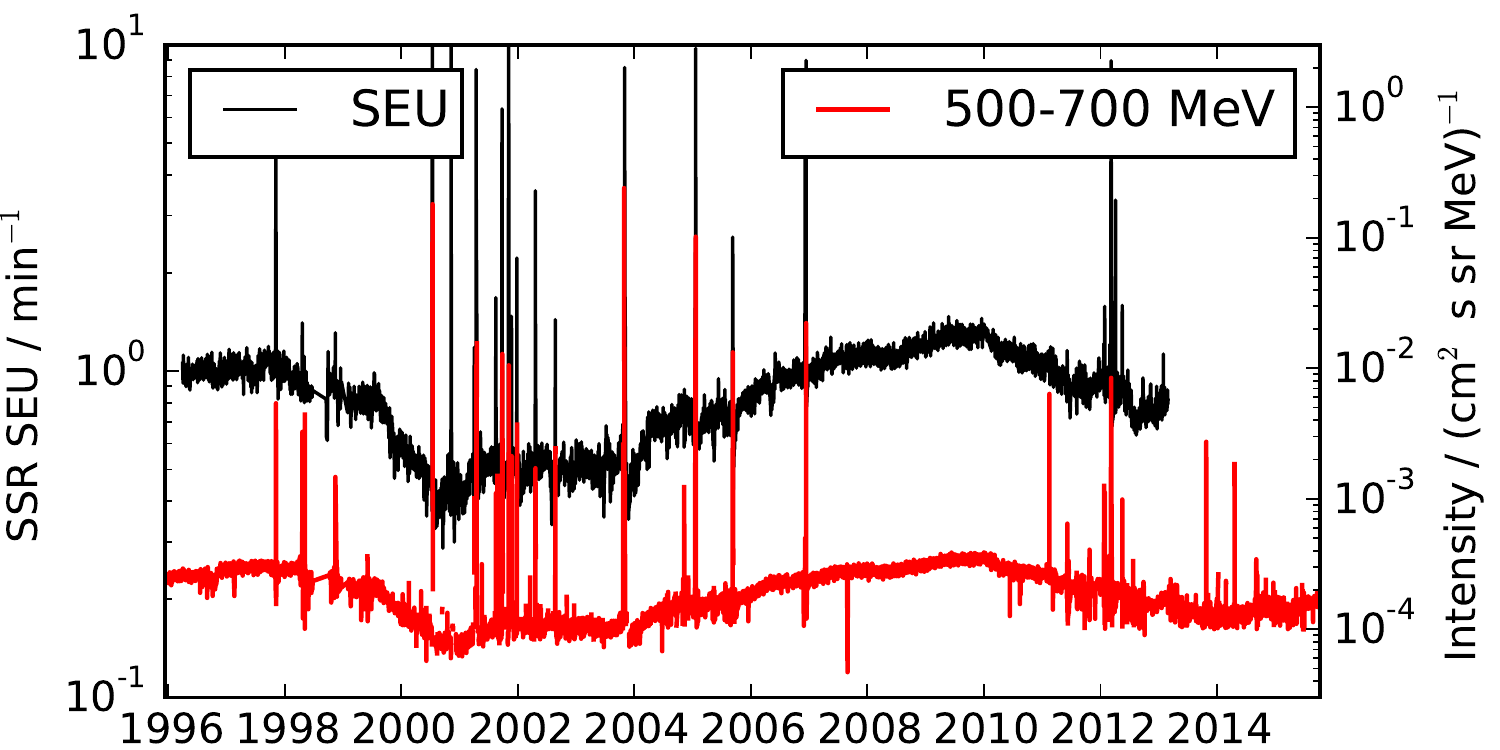}
    \caption{Single Event upsets recorded by SOHO’s solid state recorder and EPHIN measured flux of 500-700 MeV protons.}
    \label{fig:seu}
\end{figure}{}
\label{sec:penetrating}
\citet{kuehl_etal_2015} have presented a method in order to extend the energy range of \ac{EPHIN} for protons up to energies of above 1~GeV based on the Level~2 \ac{PHA} data. Utilizing extensive \ac{GEANT4} \citep{geant4} simulations of the instrument, it has been shown that the initial energy of a given proton penetrating the instrument (i.e. triggering detectors A to F) can be derived from the energy losses in \ac{SSD} C and D. Due to too high noise in detector D (ultimately leading to the switch on of the EPHIN failure mode D), the method is restricted to data before 2017. However, those high energy measurements of \acp{GCR} \citep{kuehl_etal_2016} as well as \acp{SEP} \citep{kuehl_etal_2017} have been proven to be not only of scientific value for the study of \ac{GCR} modulation as well as \ac{SEP} acceleration and transport, but can be also utilized for space weather applications. As an example, Figure \ref{fig:seu} presents the flux of 500-700~MeV protons measured by EPHIN in comparison to the single event upsets (SEUs) recorded by SOHO's solid state recorder \citep[see][for details]{Curdt_Fleck_2015}. The dependency of the SEU rate on the solar cycle variations due to the \ac{GCR} modulation as well as sporadic increases due to \ac{SEP} events can be clearly seen in this comparison. Annual averaged proton flux spectra from 250~MeV to 1.6~GeV based on EPHIN measurements can be downloaded from the cosmic ray database \citep{maurin_etal_2014}: \url{https://lpsc.in2p3.fr/cosmic-rays-db/}.

\section{Space Weather Applications}
\label{sec:spaceweatherapplications}
\subsection{Long-term Data / Space Weather Monitoring}
EPHIN data has been used in various \ac{SEP} event analysis \citep[e.g. ][]{cohen_etal_2018,dresing_etal_2018,gomez-herrero_etal_2002,lario_etal_2014} and has contributed data to multiple \ac{SEP} event catalogues \citep{sepserver,paassilta_etal_2017}. \newline
Furthermore, the availability of EPHIN data over more than two solar cycles also allows for long-term studies of the variation of the \ac{GCR} flux over two solar cycles of different magnetic polarity with special emphasis on the contribution of \ac{SEP} events to the overall fluence observed during this long-term mission. \citet{kuehl_etal_2019} has shown that for protons and helium particles the fluence during as little as 10 days contributes between 24\% and 79\% to the overall fluence during 23 years of measurements depending on the particles energy. The large \ac{SEP} events responsible for these contributions have been also linked to significant solar array degradation and SEUs in SOHO's electronics \citep{Curdt_Fleck_2015}. \newline
The derived flux for high energetic penetrating particles has been also used for validation purposes of the GLE forecast scheme HESPERIA UMASEP-500 \citep{umasep}.

\subsection{Near Real Time Data / Space Weather Forecast}
\label{sec:nearrealtime}
While most EPHIN data sets are only available with a couple of days of delay, some near real time Level~1 data are available and includes SCI and \ac{PHA} files for each available minute individually (in contrast to the daily merged files of all other data products). However, due to limited downlink possibilities, these near real time data are only available for a couple of hours each day depending on the ground station coverage \citep{malandraki_2018} . The data is provided here:
\url{http://ulysses.physik.uni-kiel.de/costep_realtime/realtime/lvl1/} \newline
\begin{figure}
    \centering
    \includegraphics[width=1\columnwidth]{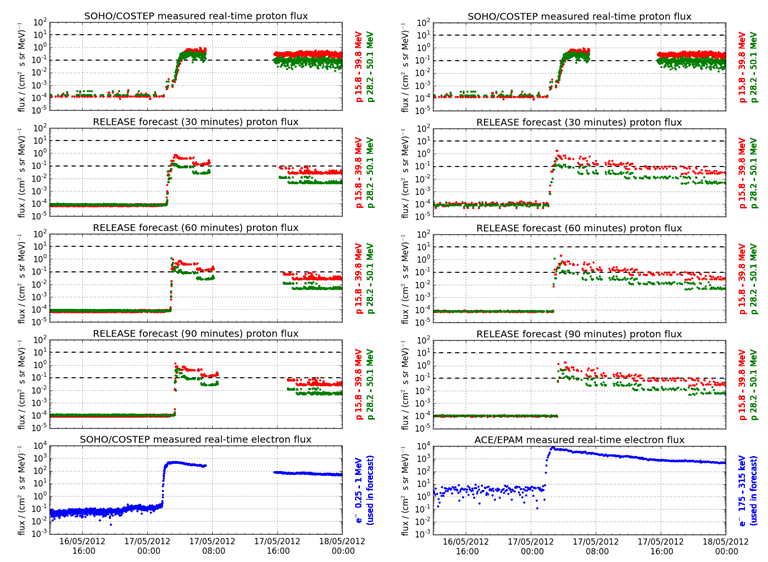}
    \caption{Performance of the HESPERIA REleASE forecast scheme based on non-realtime data from EPHIN (left) and ACE/EPAM (right) for GLE 71 in May, 2012.}
    \label{fig:release}
\end{figure}{}
One of the space weather related applications of the near real time data of EPHIN is the forecasting scheme Relativistic Electron Alert System for Exploration (REleASE) introduced by \citet{posner_2007}. \ac{SEP} events can cause an increase of the particle flux by several orders of magnitude for various particle types such as electron, protons and ions. While the radiation hazard caused by these particle populations has a strong dependency on the event characteristics such as the energy spectra and abundances of the various particle types as well as the location of the observer and the surrounding shielding, it can be generally assumed that the relativistic, minimum-ionizing electrons have a significant lower impact than the protons and ions. However, since electrons and protons are often accelerated and injected into the interplanetary medium at the same time during an \ac{SEP} event, the shorter propagation time of the electrons can be exploited in order to forecast the proton that arrive at L1 with an delay of approximately 30-90~minutes depending on the interplanetary propagation as well as the particle energy. In detail, the REleASE system utilized an empirical relationship between the expected proton flux at energies $>$30-50~MeV measured 30, 60 or 90~minutes later than the electrons dependent on measured electron flux and its increase that have been derived based on historical EPHIN data. The derived forecasting matrix can then be applied to near real time electron measurements in order to issue warnings and forecasts of the proton flux at 1AU. \citet{posner_strauss_2020} have also studied the feasibility of a REleASE system supplemented with data from a second instrument near Mars in order to provide warnings for a possible future human mission to Mars. \newline
The REleASE scheme has been recently further improved as part of the EU HORIZON 2020 HESPERIA (High Energy Solar Particle Events forecasting and Analysis) project \citep{malandraki_2018}. Besides software updates and performance studies based on historical data that have not been used in the calculation of the forecasting matrix, the HESPERIA REleASE scheme has been developed such that its innovation is to enable derivation and provision of the 30-50~MeV proton forecast based on electron measurements from ACE/EPAM \citep{epam} as well as EPHIN. The advantage of ACE/EPAM is the higher time coverage of the near real time data. HESPERIA REleASE provides forecasts in a complementary way based on both instruments in parallel. As an example of the forecast performance, Figure \ref{fig:release} shows the May, 2012 \ac{SEP} event and the related forecast based on historical data for EPHIN (left) and EPAM (right). The upper panel presents the proton flux measured by EPHIN while the three central panels show the proton forecast for 30, 60 and 90~minutes and the bottom panels shows the electron flux used for the forecast. From the figure it is clear that HESPERIA REleASE based on either of the two instruments was able to forecast the event. \newline
The real time plot as well as a collection of historical plots can be found on the HESPERIA website:
\url{https://www.hesperia.astro.noa.gr/index.php/results/real-time-prediction-tools/release}\newline
It is noteworthy that the HESPERIA REleASE model has been selected on international level and participates in the Integrated Solar Energetic Proton Alert/Warning System (ISEP) (https://ccmc.gsfc.nasa.gov/isep/). ISEP is a 3-year collaboration project between the Space Radiation Analysis Group (SRAG) at Johnson Space Center and the Community Coordinated Modeling Center (CCMC) at Goddard Space Flight Center to bring state-of-the-art space weather models from research and development at universities and small businesses to operational use at NASA (R2O). These models have a user interface in the form of the SEP scoreboard (https://ccmc.gsfc.nasa.gov/challenges/sep.php) that allows the SRAG console operator to view and compare the results from several different models simultaneously. HESPERIA REleASE has been recently successfully integrated in the CCMC/SEP scoreboard, providing unique space weather predictions for crew protection in human spaceflight in the exo-LEO mission era.

\section{Lessons Learned \& Possible Improvements}   
\label{sec:lessonslearned}
\begin{figure}
    \centering
    \includegraphics[width=1\columnwidth]{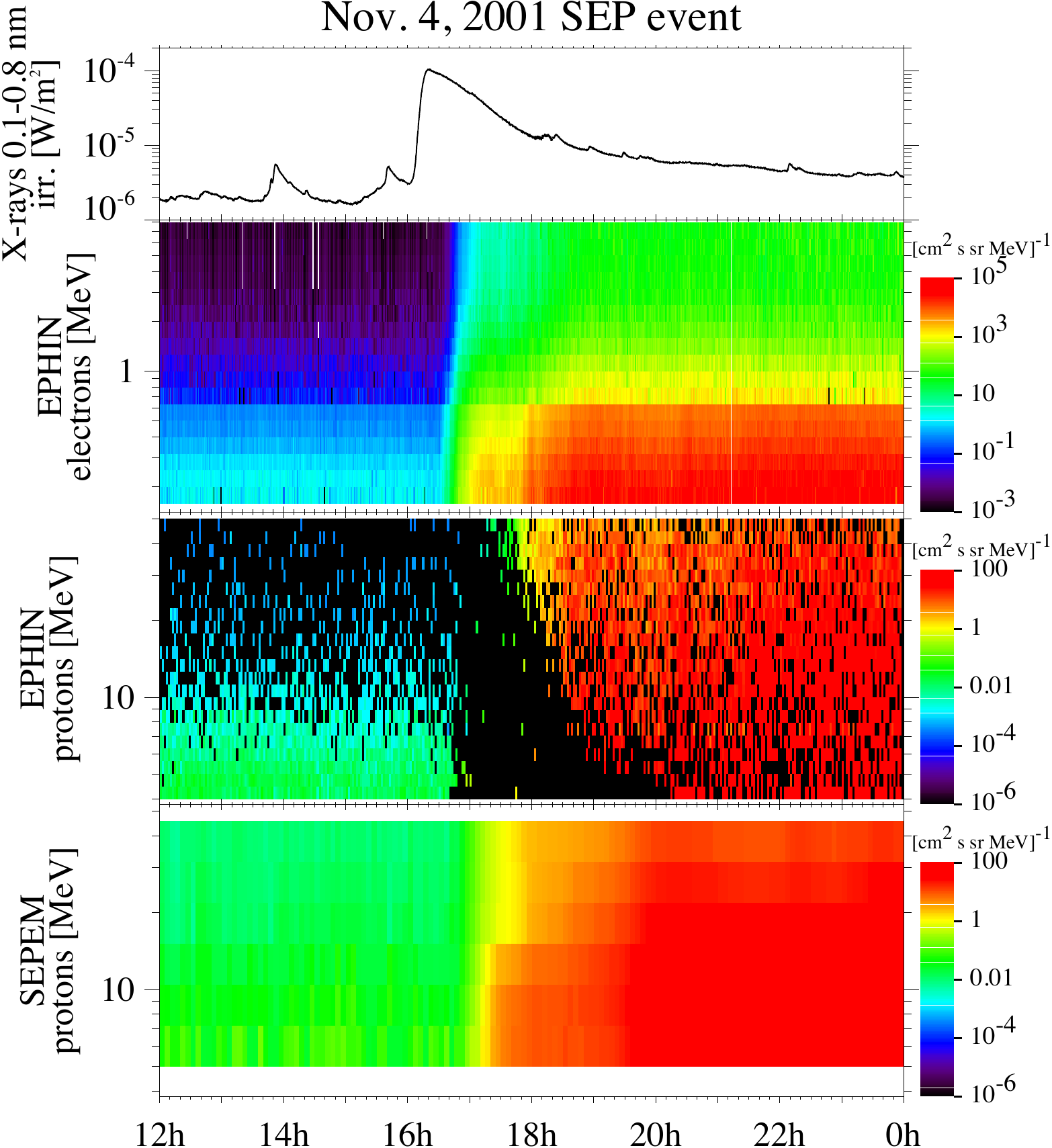}
    \caption{Solar energetic particle event onset, observed over 12 hours on Nov. 4, 2001 by GOES and SOHO. The top panel shows the GOES/SXT low-energy X-ray flux. The second panel from top shows a spectrogram of energetic electrons in the 0.25-10 MeV range observed by SOHO/COSTEP-EPHIN. Below that is displayed a 4-50 MeV proton spectrograms observed by EPHIN, and at the bottom the SEPEM energy channels of GOES-8/SEM displayed as a spectrogram. Only energy channels with full overlap between GOES 8/SEM and EPHIN are shown.}
    \label{fig:ephingoes}
\end{figure}{}
The well thought-out and robust instrument design as well as the carefully crafted data products have resulted in EPHIN being a valuable asset for particle detection not only for the space science but also for the space weather community. Its large geometric factor and low background level compared to other instruments \citep[e.g. GOES/EPS, see fig. 5 from][]{kuehl_etal_2019} allow for studies of even small \ac{SEP} events, additional data products such as the single counters have further extended the measurement capabilities and more than a billion particles evaluated and stored as \ac{PHA} data allow for detailed studies over the 25 years of measurements. However, there are also some lessons learned and possible improvements for EPHIN that should be taken into account during the design of the next generation of particle instruments. \newline
The extreme mixed radiation field of relativistic electrons, protons and heavy ions, X- and gamma rays, and secondary particles generated from the interaction of primary particles with matter pose significant challenges to energetic particle detectors such as EPHIN. This problem is illustrated in Figure \ref{fig:ephingoes}, which shows spectrograms of energetic particles measured with SOHO/COSTEP-EPHIN and GOES-8/SEM, and GOES X ray observations originating from the associated active region flare. Issues affecting EPHIN and SEM have already described for this event in the Appendix of \citet{posner_2007}. SEM essentially measures integral fluxes of energetic particles in a broad energy range, spanning from a few MeV to almost 1 GeV. It uses passive shielding and a crude way of identifying counts for the various integral channel (such as $>$10 MeV). Unfortunately, the passive shielding does not stop protons exceeding ~60 MeV from entering the detector. Multiple teams have attempted cleaning up the GOES measurements over time. Zwickl \citep[unpublished; see Appendix A. in][for details]{rodriguez_etal_2017} created differential channels by using a subtraction scheme of different integral channels, but admitted that difficulties arise that the 15--40 MeV proton channel sees with a large geometric factor ions that penetrate its passive shielding. The Zwickl-derived GOES 8/SEM for this event are shown in Figure 1 of \citet{posner_2007}. As compared to the cleaner EPHIN data set, which clearly shows that the proton event starts with an energy dispersion signature, the GOES-8/SEM onset in the 15-40 MeV channel is off by ~30 min.\newline
A more recent development is the SEPEM \citep[\url{sepem.eu},][]{jiggens_etal_2012,sandberg_etal_2014} treatment of GOES energetic proton fluxes. The output of SEPEM for the Nov. 4, 2001 event is shown in the bottom panel of Figure \ref{fig:ephingoes} as a spectrogram consisting of all differential proton channels that have full overlap with the EPHIN proton energy range. It is apparent that the SEPEM-generated onset of this event contains a large amount of high-energy protons that spill over into the 5-40 MeV proton channels, completely obscuring the actual onset of the new event. Moreover, the SEPEM spectra during this time frame suggest a decreasing intensity with energy in this energy range, which would contradict the cleaner measurement made by EPHIN. The superior performance of EPHIN during this critical onset time is thanks to the active anti-coincidence system, an essential component of any accurate SEP detection system that is supposed to provide reliable measurements while being exposed to any given particle environment. At the same time, the suppression of pre-event counts is due to the mixed radiation field EPHIN is exposed to, particularly the onset of the intense electron event. As mentioned in Section \ref{sec:coincidence}, the threshold of the coincidence system responsible for the initial particle identification scheme is set too low, allowing a large number of electrons being recorded in the low-energy proton channels. This combined with the low count rate of actual protons from the previous SEP event suppresses the number of pulse-height analyzed protons in this channel and replaces them with electrons, which are sorted out by the subsequent dE/dx~vs~E analysis. Only very few actual proton counts remain until the new SEP proton event kicks in, generating the well-known energy dispersion pattern of fast protons arriving before lower-energy slower protons. Any vertical column in the EPHIN spectrogram is representative of a momentary (2-min) spectrum of the event. It is clearly seen that the spectra are anomalous, i.e., the spectral intensity increases with energy below the dispersion onset, and above that rolls over into a nominal proton spectrum that decreases in intensity with energy. In a new detector system, this effect can effectively be reduced by adjusting the lower-energy proton channel threshold or by utilizing increased on-board processing capabilities that allow for more complicated particle separation rather than just one threshold of one detector, which would significantly reduce the spill-over of electrons into the proton channels. None of the energy dispersion phenomena are captured in the GOES 8/SEM data set shown here as part of the SEPEM output.\newline
While the dynamic range of EPHIN achieved with the automatic switch-off of the outer segments of detectors A and B during periods of high fluxes (resulting in a decrease of the geometric factor by a factor $\approx$~24) is superior compared to other instruments \citep[e.g. SOHO/ERNE, see fig. 7 from][]{kuehl_etal_2019}, it does not prevent EPHIN from suffering pile-up and dead-time issues during the largest \ac{SEP} events. Hence, a reduction of the size of the inner segments and implementing a sophisticated livetime counter in the electronics of EPHIN would have improved the data quality during these events. Furthermore, additional material surrounding the anti-coincidence detector G would have decreased the high count rates of this detector dominated by low energy particles during \ac{SEP} events and thus further reducing the dead-time of the instrument. It has to be also noted, that the ring switch-off can introduce additional uncertainties regarding the particle identification. While the ring is switched off, particles penetrating an outer segment of detector A and hitting the central segment of B at the same time as a valid particle hitting both central segments will add its energy losses in B and deeper detectors to the measurement, artificially increasing the measured energy of the detected particle and therefore resulting in a too hard spectrum. This effect could explain the hard electron spectrum in figure \ref{fig:ephingoes}, while the sudden decrease in flux above 600~keV is most likely due to an underestimation of the dead layer of detector C. Therefore, it should also be considered for future instruments to add a complimentary high flux instrument, also based on EPHIN principles, with a lower geometric factor and sufficient passive and active anti-coincidence to allow measurements of particle fluxes during the most extreme \acp{SEP} events. While this second instrument would not be able to detect small \ac{SEP} events, the combination of them would increase the dynamic range of the observable fluxes. The realization of such a design can be inspired by the pixel detector used for the PSP/IS\sun IS/EPI-Hi \citep{mccomas_etal_2016}. The possibility of using two or more instruments with adjustable geometric factor similar to EPHIN, each optimized for measuring only in a small energy range should be also explored as an alternative approach. \newline
\ac{GEANT4} simulations show that $\delta$-electrons generated by ions stopping in detector A can propagate to detector B and cause energy depositions above the threshold of 60~keV. This issue of particles below the nominal energy range being counted by the coincidence logic could have been easily resolved by increasing the threshold to above 100~keV.\newline
It is moreover recognized that proton energies of 30-100 MeV are of significant relevance for radiation safety, as they penetrate deeper in matter and thus can more readily reach the inside of exploration vehicles as well as astronauts in extravehicular suits than protons of lower energies. The EPHIN limitation of stopping protons of $<$54 MeV could have been mitigated by either using more and/or thicker detectors at the rear of the particle telescope or by including a scintillation detector in the detector stack. The former solution provides better energy resolution on the costs of requiring more channels or high bias voltage for thicker detectors. These thicker detectors, such as the lithium drifted ones used for EPHIN, are also less reliable for a long-term mission (see section \ref{sec:ephin}).  \citet{mccomas_etal_2016}, however, have shown that two successive detectors connected to the same front-end electronics as used for the PSP/IS\sun IS/EPI-Hi instrument can mitigate these downsides. Utilizing a scintillation detector as used for i.e. the Energetic Particle Detector on Solar Orbiter \citep[][]{solo-epd} provides a high stopping power especially for heavier ions without the need of several channels while reducing the energy resolution and introducing systematic uncertainties due to higher temperature dependencies as well as non-linearities such as quenching.\newline
Histograms that store the count rate distribution for protons, Helium but also for energetic electrons in all 36 virtual telescopes (cf. section \ref{sec:lvl2pha}) would have allowed to compute the directional distribution without the statistical limitation of an \ac{PHA} based analysis.\newline
The calculation of the high energy spectra of penetrating particles is limited by statistics since it is based on \ac{PHA} data. This limitation could have been eliminated by providing an energy loss histogram for particles penetrating the entire detector stack in the histogram data. However, the method inhabits a significant systematic uncertainty due to EPHIN's inability of separating forward and backward penetrating particles as well as relativistic protons and electrons \citep{kuehl_etal_2016}. The E6 Instruments on the Helios spacecrafts have shown that adding a Cherenkov detector behind the detector stack can be utilized in order to distinguish between both directions \citep{marquardt_heber_2019} while the COSPIN/KET instrument on Ulysses \citep{cospin} has proven the possibility to distinguish between relativistic protons up to 2.1~GeV and electrons utilizing an Aerogel Cherenkov detector.\newline
An energy-loss histogram of one of the inner detectors in anti-coincidence to all other detectors would have provided more capabilities in the analysis of $\gamma$-rays following the approach of the Lunar Lander Neutron and Dosimetry experiment (LND) on Chang'e4 \citep{lnd} and PSP/IS\sun IS/EPI-Hi \citep{mccomas_etal_2016}.  \newline
By utilizing the increased on-board processing capabilities of these state of the art instruments compared to the 25 year old EPHIN electronics, the proposed improvements could be included without the need of significant higher telemetry rate. For example, the 2.8 Gbit per orbit telemetry from PSP/IS\sun IS/EPI-Hi translates to an average of 192 bit per second for the first orbit of 168 days which is similar to EPHINs nominal telemetry rate of 172 bit per second. The mass and power budget of such an improved instrument, however, would depend on the detailed design and is therefore difficult to estimate.

\section{Summary and Outlook}
Providing 25 years of continuous measurements during two entire solar cycles and a variety of different solar activities resulting in different \ac{SEP} environments, EPHIN on-board SOHO is a valuable asset regarding particle detection for the space weather and solar science communities. Due to the low background thanks to the lateral anti-coincidence and the large geometric factor as well as the availability of PHA data (including penetrating particles), it was possible to provide new data products with enhanced scientific value utilizing revised simulations (GEANT4) of the already 25 year old instrument. The numerous data products summarized here have been used for multiple long-term cosmic ray and \ac{SEP} studies as well as space weather applications such as \ac{SEP} event forecast and the analysis of single event upsets in spacecraft electronics.  \newline
On the technical level, a number of benefits and limitations of EPHIN have been pointed out, which a future instrument best qualified for space weather purposes should incorporate and overcome, respectively.\newline
In order to accurately evaluate SEP forecasting systems, it is imperative that the actual onsets of SEP events can be determined, both in terms of actual event onset by differential proton channel, and in terms of when a differential proton channel exceeds a threshold level that is relevant for protecting human explorers in deep space. EPHIN is currently one of only few detectors that have this capability due to the large dynamic range (which could be further improved by adding a second high flux instrument) as well as the low background and spill-over of different particle types mainly accomplished by the anticoincidence system. Furthermore, a variety of data products that have extended the measurement capabilities of EPHIN have been derived and provided to the community based on large amount of \ac{PHA} data.\newline
Improvements would include a slightly higher energy range (up to $\approx$100~MeV) achieved by adding more SSDs, a complex and sophisticated onboard particle separation as well as further passive material surrounding the active parts in order to decrease pile up effects and deadtime limitations. The possibility of an enhanced detection of neutral particles should also be taken into account.\newline
We propose a Proton Electron Comprehensive Telescope for Moon/Mars Exploration (ProtECT-ME) based on the EPHIN design that would incorporate these features and improvements in order to be best suited for radiation monitoring and providing real time data for \ac{SEP} forecast schemes.\newline
With further improvements such as a rejection system for backward penetrating particles and an improved separation between relativistic electrons and protons as well as onboard data products such as histograms for penetrating particles, such an instrument could be also used for aviation \ac{SEP} warnings.

\acknowledgements

{The SOHO/EPHIN project is supported under grant 50 OC 1702 by the German Bundesministerium f\"ur Wirtschaft through the Deutsches Zentrum f\"ur Luft- und Raumfahrt (DLR). This project has received funding from the European Union’s Horizon 2020 research and innovation program under grant agreement No. 637324. RGH acknowledges the financial support by the Spanish Ministerio de Ciencia, Innovaci\'on y Universidades FEDER/MCIU/AEI Projects ESP2017-88436-R and PID2019-104863RB-I00.}


\bibliographystyle{swsc}
\bibliography{biblio}

\end{document}